\documentstyle[12pt,epsf]{article}
\newcommand{\abstracts}[1]{{
\centering{\begin{minipage}{12.2truecm}
\normalsize\baselineskip=15pt
\centerline{\footnotesize ABSTRACT}\vspace*{0.3cm}
\parindent=20pt #1
\end{minipage}}\par}}
\newcommand{\dual}{\mbox{}^{\ast}}
\newcommand{\dd}{\mbox{\rm d}}

\newcommand{\beqn}{\begin{eqnarray}}
\newcommand{\eeqn}{\end{eqnarray}}
\newcommand{\eq}[1]{(\ref{#1})}
\newcommand{\beq}{\begin{equation}}
\newcommand{\eeq}{\end{equation}}

\begin{document}
~\vspace{-1.5cm}
\begin{flushright}
{\large ITEP-TH-69/97}
\end{flushright}
\vspace{1cm}

\begin{center}

{\baselineskip=16pt
{\Large \bf Vortex Dynamics in Classical Non--Abelian Spin
\vspace{2mm}

Models}\\

\vspace{1cm}

{\large O.A.~Borisenko
\footnote{email: oleg@ap3.gluk.apc.org}$^{\! ,a}$ ,
M.N.~Chernodub\footnote{email:
chernodub@vxitep.itep.ru}$^{\! ,b}$ and F.V.~Gubarev\footnote{email:
gubarev@vxitep.itep.ru}$^{\! ,b}$}\\

\vspace{.5cm}
{ \it

$^a$ N.N.Bogolyubov Institute for Theoretical Physics, National\\ Academy
of Sciences of Ukraine, 252143 Kiev, Ukraine

\vspace{0.3cm}

$^b$ ITEP, B.Cheremushkinskaya 25, Moscow, 117259, Russia

}
}
\end{center}

\vspace{1.5cm}

\abstracts{
We discuss the abelian vortex dynamics in the
abelian projection approach to non-abelian spin models.
We show numerically that in the three-dimensional
$SU(2)$ spin model in the Maximal Abelian projection
the abelian off-diagonal vortices
are not responsible for the phase transition
contrary to the diagonal vortices.
A generalization of the abelian projection approach
to $SU(N)$ spin models is briefly discussed.}

\vspace{1cm}

%\newpage

%%% For alphabetic footnotes indices in text  %%%%
\setcounter{footnote}{0}
\renewcommand{\thefootnote}{\alph{footnote}}
%%%%%%%%%%%%%%%%%%%%%%%%%%%%%%%%%%%%%%%%%%%%%%%%%%

\baselineskip=14pt

\section{Introduction}

One of the important phenomena in spin models is the
mass gap generation. This phenomena have intensively
been studied for the abelian spin models in various space dimensions.
Classical papers by Berezinskii, Kosterlitz and Thouless \cite{BKT} show
that in $2D$ $XY$ model the mass gap generation is due to
the condensation of topological excitations called vortices.
In a massive phase the vortices are condensed and this leads to the
exponential falloff of the two-point correlation function
while at the weak coupling the vortices form a dilute gas
with the logarithmic long-range interaction (see \cite{fr}
for a rigorous proof).
In the three-dimensional $XY$ model the condensation of vortices
also leads to the mass gap generation \cite{KoShWi86,PoPoYu91}.
In this case the situation is somewhat simpler since one has
a spontaneous breaking of the global $U(1)$ symmetry at the weak
coupling phase and vortices interact via the short-range Coulomb
potential above the critical point.

The nature of mass gap in non-abelian
spin systems is still an open question despite numerous
attempts to attack this problem \cite{NatureMassGap}.
The non-abelian models, similarly to their gauge partners, possess
a number of excitations (instantons, merons, thin and thick vortices)
and this is a delicate problem to estimate their contribution
to the mass gap reliably. Nevertheless, one can handle the problem,
at least partially, using the gauge fixing procedure and calculating
the contribution to the mass gap from configurations which the gauge
fixing leaves untouched. This idea came obviously up from lattice gauge
models where it was successfully applied for studying the confinement
mechanism. We put forward recently such an approach in
Ref.~\cite{BoChGu97} making use the abelian projection
method~\cite{tH81}.  After abelian projection the non-abelian spin
system possesses abelian symmetries and abelian topological excitations
(vortices). It is important to stress at this point that, precisely
like in gauge models, the abelian projection does not mean the restriction
of the configuration space of the original non-abelian model. The original
non-abelian global symmetry is broken up to its maximal abelian
subgroup by appropriate gauge fixing procedure.  In general, global
gauge fixing procedure is a powerful tool for investigating
nonabelian models in different regimes. Also it was understood pretty
long ago that this is a necessary condition for construction of the
correct perturbation theory of non-abelian models in the weak
coupling region \cite{Hasenfr}.  We would like to argue that this
method is also very fruitful in studying the nonperturbative
phenomena like the mass gap generation.

Following this avenue, we have found the abelian
dominance~\cite{SuYo90} in the Maximal Abelian (MaA)
projection\footnote{The MaA projection was first used for study of
the lattice non-abelian gauge theories in Ref.~\cite{KrLaScWi87}.} of
the $SU(2)$ spin model: the full mass gap (i.e., calculated from full
correlation function) in classical non-abelian $SU(2)$ spin model is
dominated by the abelian mass gap calculated from the projected
correlation function \cite{BoChGu97}. We have concluded that the
abelian degrees of freedom seem to play an important role in the mass
gap generation phenomena.

The next important question is what is the role of abelian
topological excitations in the phenomena of the abelian dominance.
The $3D$ $SU(2)$ spin model possesses two types (``diagonal'' and
``off-diagonal'') of the abelian vortices.  The diagonal vortices
were shown to be condensed in the massive phase and they form the
dilute gas in the massless phase~\cite{BoChGu97}. This picture is
very similar to the vortex dynamics in the abelian $XY$ model.  In
this paper we study the behavior of the off-diagonal vortices across
the phase transition. The generalization of the abelian projection
approach to the $SU(N)$ spin models is discussed at the end of the
paper.

\section{Off-Diagonal Vortices in $SU(2)$ Spin Model}

We study the three-dimensional
$SU(2)$ spin model with the following action
\beqn
S = - \frac{\beta}{2} \sum_x \sum^3_{\mu=1} {\rm Tr} \, U_x \,
    U^+_{x+\mu} \,, \label{actionSU2}
\eeqn
where $U_x$ are the spin fields taking values in the $SU(2)$ group
and $\beta$ is the coupling constant.  Action \eq{actionSU2} is
invariant under $SU_L(2) \times SU_R(2)$ global transformations, $U_x
\to U^{(\Omega)}_x = \Omega^+_L \, U_x \, \Omega_R$, where
$\Omega_{L,R}$ are the $SU(2)$ matrices.

The Maximal Abelian projection for the $SU(2)$ spin theory
is defined by the following maximization condition~\cite{BoChGu97}:
\beqn
  \max_{\{ \Omega \}} R [U^{(\Omega)}]\,,\qquad
  R [U] = \sum_x {\rm Tr} \Bigl( U_x \sigma^3 U^+_x \sigma^3 \Bigr)\,.
  \label{R}
\eeqn
The functional $R [U]$ is invariant under $U_L(1) \times U_R (1)$
global transformations, $U_x \to U^{({\tilde \Omega})}_x =
{\tilde \Omega}^+_L U_x {\tilde \Omega}_R$, where
${\tilde \Omega_{L,R}} = e^{i \sigma^3 \omega_{L,R}}$ and
$\omega_{L,R} \in [0, 2 \pi)$. Due to the invariance of the
functional $R$ under the abelian gauge transformations,
the condition \eq{R} fixes the $SU_L(2) \times SU_R (2)$
global symmetry group up to $U_L(1) \times U_R (1) \sim O_L (2)
\times O_R (2)$ global group.

In order to fix the MaA gauge in the path integral approach we
substitute the Faddeev--Popov (FP) unity
\beqn
1 = \Delta_{FP}[U;\lambda ] \int D \Omega \exp\{ \lambda
R[U]\}\,, \ \  D \Omega = D\Omega_L \, D\Omega_R
\label{star2}
\eeqn
into the partition function of $SU(2)$ spin model. Here $\Delta_{FP}$ is the FP
determinant and the limit $\lambda \to + \infty$ is assumed.
The functional $R[U]$ is defined in (\ref{R}).
Shifting the fields $U_x \to U^{\Omega^+}_x$ and using the invariance
of the FP determinant $\Delta_{FP}[U;\lambda ]$, the action $S[U]$ and
the Haar measure $D U$, we get the product of the gauge orbit volume
$\int D\Omega$ and the partition function with the fixed gauge:
\beqn
Z_{MaA} = \int D U \exp\{- S[U] + \lambda R[U] \}
\Delta_{FP}[U;\lambda ]\,.
\label{PFFG}
\eeqn
The limit $\lambda\to\infty$ should guarantee that all the saddle points of
invariant integral in (\ref{PFFG}) lie in the abelian subgroup.

Let us parametrize the $SU(2)$ spin field $U$ in the standard way:
$U^{11}_x = \cos \varphi_x e^{i\theta_x}; $
$\ U^{12}_x = \sin
\varphi_x e^{i\chi_x}; \ U^{22}_x = U^{11, *}_x; \
U^{21}_x = - U^{12,*}_x; \ 0 \le \varphi \le \pi/2, \ 0 <
\theta,\chi \le 2 \pi$. Under the $U_L(1) \times U_R (1)$ transformations
defined above, components of the field $U$ transform as
\beqn
 \theta_x \to \theta^\prime_x
 = \theta_x + \omega_d \quad {\rm mod}\, 2 \pi\,,\qquad
 \chi_x \to \chi^\prime_x & =
 \chi_x + \omega_o \quad {\rm mod}\, 2 \pi\,,
\eeqn
where $\omega_{d,o} = - \omega_L \mp \omega_R $. It is
convenient to decompose the residual symmetry group,
$O_L (2) \times O_R (2) \sim O_d (2) \times O_o (2)$. The diagonal
(off-diagonal) component $\theta$ ($\chi$) of the $SU(2)$ spin $U$
transforms as a spin variable with respect to the $O_d(2)$ ($O_o(2)$)
symmetry group.

It is useful to consider the one-link spin action $S_l$
in terms of the angles $\varphi$, $\theta$, $\chi$:
\beqn
S_{l_{x,\mu}}
= - \beta \Bigl[\cos\varphi_x \cos\varphi_{x+\hat\mu}
\cos(\theta_x - \theta_{x+\hat\mu})
+ \sin\varphi_x \sin\varphi_{x+\hat\mu}
\cos(\chi_x - \chi_{x+\hat\mu})
\Bigr]\,.
\label{SU2Act}
\eeqn
The action consists of two parts which correspond to the
self--interaction of the spins $\theta$ and $\chi$, respectively. The
$SU(2)$ component $\varphi$ does not behave as a spin field and its
role is to provide the interaction between the $\theta$ and $\chi$
spins. Thus, the $SU(2)$ spin model in the abelian projection reduces
to two interacting copies of the $XY$ model with the fluctuating
couplings due to the dynamics of the field $\varphi$.

The phases $\theta= \arg U^{11}$ and $\chi = \arg U^{12}$
behave as the $O(2)$ spins under the abelian gauge
transformation~\cite{BoChGu97}. In the abelian projection
the $SU(2)$ spin model possesses two types of the abelian vortices due
to the compactness of the residual abelian
group. These vortices correspond to the diagonal ($\theta$) and to the
off-diagonal ($\chi$) abelian spins (``diagonal'' and ``off-diagonal''
vortices, respectively).

We expect that in the MaA projection the diagonal
vortices may be more dynamically important then
the off-diagonal vortices. The reason for this expectation is simple.
In our representation for $SU(2)$ spin field
the maximizing functional \eq{R} has the form:
$R = 4 \sum_x \cos^2 \varphi_x + const$.
Therefore, in the MaA projection the effective coupling constant
in front of the action for the $\theta$ spins is maximized while
the effective self--coupling for $\chi$ spins
is minimized. Thus we may expect that the diagonal vortices
may be more relevant to the dynamics of the
system than the off-diagonal vortices.

The behavior of the diagonal vortices was studied numerically
in Ref.~\cite{BoChGu97}. The diagonal abelian vortices are
condensed in the massive phase and they form the dilute
gas of the vortex anti--vortex pairs in the Coulomb phase.
This behavior is very
similar to the behavior of the abelian vortices in $3D$ $XY$
model~\cite{KoShWi86,PoPoYu91},
where the vortices are known to be responsible for the mass gap generation.
This similarity allows to conclude that the diagonal abelian vortices in
the MaA projection of $SU(2)$ spin model are relevant degrees of freedom
for the mass gap generation~\cite{BoChGu97}.

Now we proceed to study the condensation properties of the
off--diagonal vortices in the MaA projection across the phase transition.
First, we construct the off-diagonal vortex trajectory $\dual j_o$ from the spin
variables $\chi$ using the standard formula \cite{BaMyKo77,Sa78}:
$\dual j_o = {(2 \pi)}^{-1} \dual \dd [\dd \chi]$,
where the square brackets stand
for ``modulo $2 \pi$'' and the operator ``d'' is the lattice
derivative. In the three dimensions the vortex
trajectories are loops which are closed due to the property $\dd^2=0$.

The condensation of the vortex trajectories can be studied by measuring
of the so-called percolation probability \cite{PoPoYu91} $C$:
\beqn
C = \lim_{r \to \infty}{\left(\sum\limits_{x,y,i}
\delta_{x\in \dual j_i} \, \delta_{y\in \dual j_i} \cdot
\delta\Bigl(|x-y|-r\Bigr)\right)} \cdot {\left(
\sum\limits_{x,y} \delta\Bigl(|x-y|-r\Bigr)\right)}^{-1}\,,
\eeqn
where the summation is over all the vortex trajectories
$j_i$, and over all the points $x$, $y$ of the lattice.

If there is a non-zero probability for two infinitely separated points
to be connected by a vortex trajectory then the quantity $C$ is not zero.
This means that the entropy of the vortex trajectories dominates over the
energy of vortices, which means in turn
that the vortex condensate exists in the
vacuum. If the vortex trajectories form an ensemble of small loops
(the gas of vortex--anti-vortex pairs) then the quantity $C$ is obviously
zero.

We studied the quantity $C$ by numerical methods on the $16^3$
lattice with periodic boundary conditions. In our numerical
simulations the Wolff cluster algorithm~\cite{Wolff} has been used.
In order to thermalize the spin fields at each value of the coupling
constant $\beta$ we performed a number of thermalization sweeps,
which is much greater than measured auto-correlation time.

We show the percolation probability $C$ for both diagonal (taken from
Ref.~\cite{BoChGu97}) and off-diagonal vortex trajectories in Fig.~1.
One can see that the percolation probability of the off--diagonal
vortex trajectories does not vanish in the massless phase contrary to
that of diagonal vortices. This allows us to conclude that the
off--diagonal vortices do not play a significant role in the mass gap
generation\footnote{The properties of the diagonal and off-diagonal
vortices in the $SU(2)$ spin model are very similar to the properties
of, respectively, the monopoles and minopoles in the MaA projection
of the $SU(2)$ gluodynamics, respectively \cite{ChPoVe95}.} and that
the only diagonal sector of the $SU(2)$ spin model in the MaA
projection may be responsible for the mass gap.

\section{The MaA projection for $SU(N)$ spin models}

The generalization of the MaA projection to the $SU(N)$ spin models
is quite straightforward. The maximization functional is
\beqn
R[U] = \sum_x \sum_h {\rm Tr} \Bigl( U_x \lambda_h U^+_x \lambda_h\Bigr)\,,
\eeqn
where the index $h$ runs over the Cartan subgroup of the $SU(N)$
group and $\lambda_a$ are the generators of the $SU(N)$ group. The
functional $R$ is invariant under $ {[U(1)]}^{N-1}_L
\times{[U(1)]}^{N-1}_R$ global symmetry group.
Gauge fixed partition function has the form (\ref{PFFG})
where one has to take the corresponding $SU(N)$ action and the functional
$R[U]$. Now, there are
$N(N+1)\slash 2 -1$ independent abelian spin fields:  $N-1$ diagonal
fields and $N(N-1)\slash 2$ off-diagonal spin fields.  Therefore
there are $(N-1)$ species of the diagonal vortices and $N(N-1)\slash
2$ species of the off-diagonal vortices, respectively. In analogy
with $SU(2)$ spin model one can expect that the phase transition in
the $SU(N)$ spin model is accompanied by the condensation of the
$N-1$ species of the diagonal vortices. The off-diagonal vortices
should show a random distribution and should not be relevant for mass
gap generation in the $SU(N)$ spin model.

\section*{Acknowledgments}

Authors are grateful to M.I.~Polikarpov for fruitful discussions.
M.Ch. and F.G. were supported by the JSPS Program on
Japan -- FSU scientists collaboration, by the Grants INTAS-94-0840,
INTAS--RFBR-95-0681, and by the grant number 93-02-03609, financed by
the Russian Foundation for the Fundamental Sciences.

\newpage

\section*{Figure}

\begin{figure}[htb]
%\vspace{-6cm}
\centerline{\epsfxsize=.95\textwidth\epsfbox{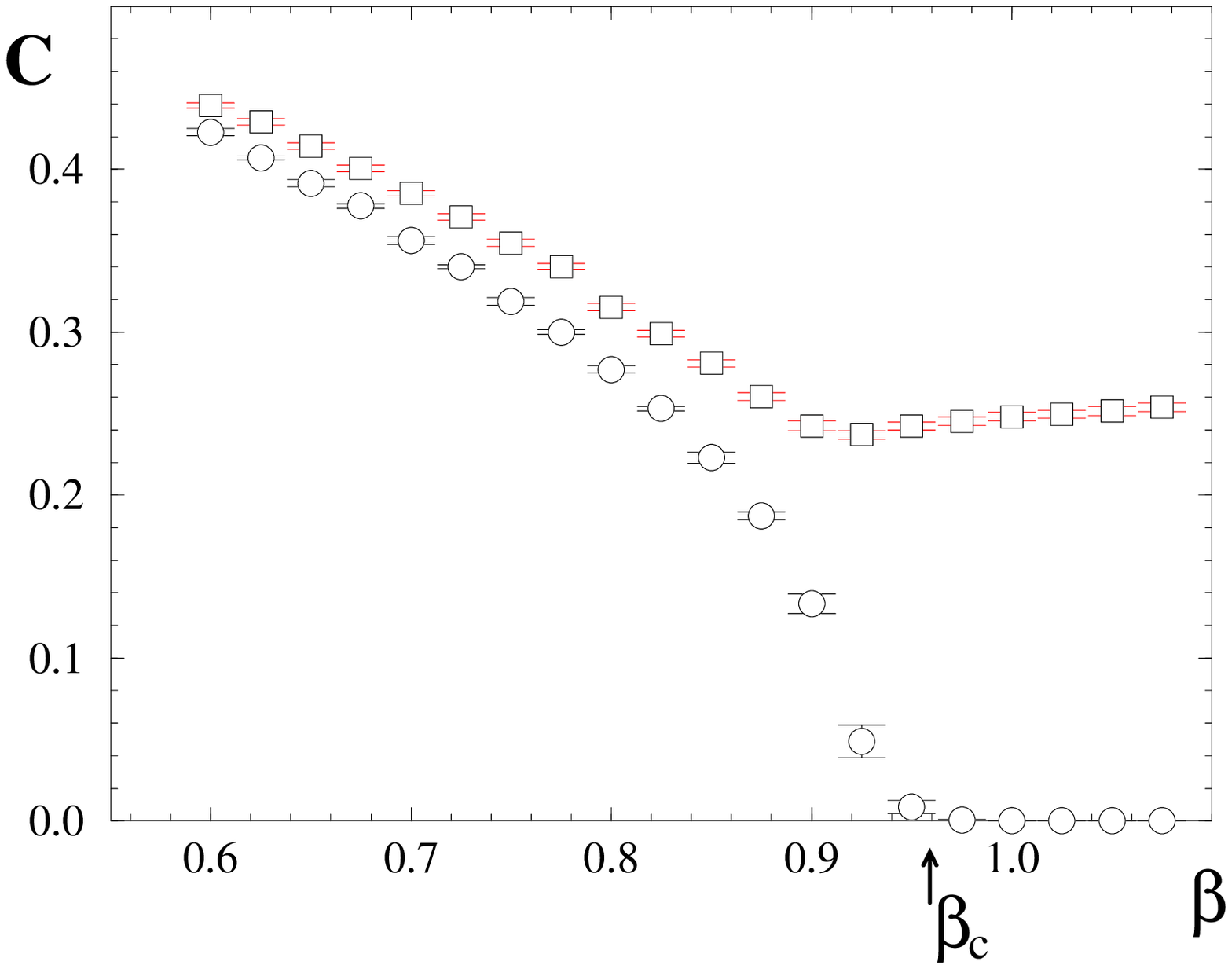}} %\eps..scale
\vspace{0.6cm}
\centerline{\parbox{15cm}{Fig.~1:
The percolation probability $C$ for the diagonal
(circles)  and anti--diagonal (squares) abelian vortex
trajectories {\it vs.} $\beta$ on the $16^3$ lattice.
The data for diagonal vortex trajectories
is taken from Ref.~\cite{BoChGu97}.}}
\end{figure}


\newpage

\begin{thebibliography}{99}

\bibitem{BKT} V.L.~Berezinskii, Sov.Journal, JETP, {\bf 32} (1971) 493;
J.M.~Kosterlitz, D.J.~Thouless, J.Phys. {\bf C6} (1973) 1181.

\bibitem{fr} J.~Fr\"ohlich, T.~Spencer, Comm.Math.Phys.
81 (1981) 527.

\bibitem{KoShWi86} G.~Kohring, R.E.~Shrock and P.~Wills,
{\it Phys.~Rev.~Lett.} {\bf 57} (1986) 1358.

\bibitem{PoPoYu91} A.V.~Pochinsky, M.I.~Polikarpov and B.N.~Yurchenko,
{\it Phys.~Lett.} {\bf A154} (1991) 194;
A.~Hulsebos, {\it preprint LTH-324}, {\tt hep-lat/9406016}; A.~Hulsebos,
{\it Nucl.~Phys.} {\bf B} {\it (Proc.~Suppl.)} {\bf 34} (1994) 695,
{\tt hep-lat/9311042}.

\bibitem{NatureMassGap} T.G.~Kov\'acs, Thesis of PHD dissertation, University
of California, Los Angeles, 1996; Nucl.Phys. {\bf B482} (1996) 613-638,
and references therein.

\bibitem{BoChGu97} O.A.~Borisenko, M.N.~Chernodub and F.V.~Gubarev,
{\it preprint KANAZAWA-97-05}, {\tt hep-lat/9705010}, to appear in
{\it Phys.Lett.}{\bf B}

\bibitem{tH81} G.~'t~Hooft, {\it Nucl.~Phys.} {\bf B 190 [FS3]} (1981) 455.

\bibitem{Hasenfr} P.~Hasenfratz, Phys.Lett. B141 (1984) 385.

\bibitem{SuYo90} T.~Suzuki and I.~Yotsuyanagi, {\it Phys.~Rev},
{\bf D42} (1990) 4257.

\bibitem{KrLaScWi87} A.S.~Kronfeld {\it et all.}, {\it Phys.~Lett.}
{\bf 198B} (1987) 516; A.S.~Kronfeld, G.~Schierholz and U.J. Wiese,
{\it Nucl.~Phys.} {\bf B293} (1987) 461.

\bibitem{BaMyKo77}
T.~Banks, R.~Myerson and J.~Kogut, {\it Nucl.~Phys.} {\bf B129} (1997) 493.

\bibitem{Sa78} R.~Savit, {\it Phys.~Rev.} {\bf B17} (1978) 1340.

\bibitem{Wolff} U.~Wolff,{\it Phys.~Rev.~Lett.} {\bf 62} (1989) 361.

\bibitem{ChPoVe95} M.N.~Chernodub, M.I.~Polikarpov and A.I.~Veselov,
{\it Phys.~Lett.} {\bf B342} (1995) 303.

\end{thebibliography}
\end{document}